# Differences in MEG and EEG power-law scaling explained by a coupling between spatial coherence and frequency: a simulation study

Bénar CG[1], Grova C[2,3,4,6], Jirsa VK[1], Lina JM[5,6,7]

[1] Aix Marseille Univ, INSERM, INS, Inst Neurosci Syst, Marseille, France

[2] Montreal Neurological Institute and Hospital, McGill University Montreal, QC, Canada;

[3] Multimodal Functional Imaging Laboratory, Biomedical Engineering Department, McGill University Montreal, QC, Canada;

[4] PERFORM Centre and Physics Department, Concordia University Montreal, QC, Canada.

[5] Département de Génie Électrique, École de Technologie Supérieure, Montréal, QC, Canada

[6] Centre de Recherches Mathématiques, Montréal, QC, Canada.

[7] Centre d'Etudes Avancées en Médecine du Sommeil, Hôpital Sacré Cœur, Montréal, QC, Canada

**Abstract (300 words)**

Electrophysiological signals (electroencephalography, EEG, and magnetoencephalography , MEG), as many natural processes, exhibit scale-invariance properties resulting in a power-law (1/f) spectrum. Interestingly, EEG and MEG differ in their slopes, which could be explained by several mechanisms, including non-resistive properties of tissues. Our goal in the present study is to estimate the impact of space/frequency structure of source signals as a putative mechanism to explain spectral scaling properties of neuroimaging signals.

We performed simulations based on the summed contribution of cortical patches with different sizes (ranging from 0.4 to 104.2 cm$^2$). Small patches were attributed signals of high frequencies, whereas large patches were associated with signals of low frequencies, on a logarithmic scale. The tested parameters included i) the space/frequency structure (range of patch sizes and frequencies) and ii) the amplitude factor *c* parametrizing the spatial scale ratios. We found that the space/frequency structure may cause differences between EEG and MEG scale-free spectra that are compatible with real data findings reported in previous studies. We also found that below a certain spatial scale, there were no more differences between EEG and MEG, suggesting a limit for the resolution of both methods.

Our work provides an explanation of experimental findings. This does not rule out other mechanisms for differences between EEG and MEG, but suggests an important role of spatio-temporal structure of neural dynamics. This can help the analysis and interpretation of power-law measures in EEG and MEG, and we believe our results can also impact computational modeling of brain dynamics, where different local connectivity structures could be used at different frequencies.

**Introduction**

Electrophysiological signals, as many other natural processes, exhibit scale-free characteristics expressed as a $1/f^\gamma$ spectrum (Freeman 2005). In practice, this implies that a spectrum is piece-wise linear in a log-log representation, the slope of which (parameter ϒ) could be an important marker of brain state (He 2014). In the case of local field potentials, several mechanisms have been proposed in order to explain these scale-free properties, as reviewed in (G. Buzsaki et al. 2012) and (Pesaran et al. 2018): low-pass frequency properties of dendrites, capacitive properties of tissues, network properties including relationships between phase coherence and distance, noise from evenly distributed ion channels across dendrites. Other putative mechanisms are more generic, taken from the domain of spatiotemporal pattern formation on the cortex (Jirsa 2009).

An interesting observation is that EEG and MEG have different 1/f profiles, with EEG having a steeper slope in a log-log representation. This was shown clearly in the work of Dehghani and colleagues, which interpreted these findings as resulting from non-resistive tissue properties (Dehghani et al. 2010). In this study, a theoretical investigation was conducted that predicts similar frequency dependence for MEG and EEG within a purely resistive[1] medium, thus suggesting an impact of non-resistive aspects of the volume conductor on the difference between slopes. The authors develop there a mathematical framework that describes the electric and magnetic field in a linear but not purely resistive framework, where the measures at sensor level depend on the frequency. This can explain why EEG and MEG are differently filtering brain activity.

An alternative - and not exclusive - hypothesis would be a dependence between the size of coherent cortical active regions and the frequency of oscillatory activity, as proposed in early works (Pfurtscheller and Cooper 1975). Indeed, the fact that EEG and MEG have different levels of spatial smearing (due to the little influence of skull on magnetic signals (Hämäläinen et al. 1993)) could result in different levels of spatial signal cancellation at different frequencies (such cancellation effects are illustrated very schematically in Figure 1). Different scenarios (resistive, non-resistive, with or without dependence of extent of active tissue on frequency) are shown in a schematic manner on Figure 2.

In order to test this latter hypothesis, within a resistive medium, we conducted a series of simulations where the "granularity" of signals at the cortical level (i.e. the size of coherent cortical activity) was varied across frequencies, with higher level of spatial coherence for low frequencies. Our objective was to show that such spatio-temporal coupling - resulting from pure geometrical considerations within a resistive medium – may result in differences in EEG and MEG spectra that are consistent with observations done on real signals.

---

[1] In a purely resistive medium, the propagation of electromagnetic fields only depends on the electrical resistance of the different components (here, brain, skull, CSF, scalp…). Importantly, in this case, there is no dependency of the observed fields on frequency. In other words, there is no filtering effect of the tissues, in contrast with non-resistive tissues where signals may be attenuated at high frequencies.

**Materials and methods**

*Cortex parcellation and EEG/MEG forward modeling*

We used the geometry for cortex, MEG sensor placement and EEG electrode localization from a patient recorded for source localization of epileptic discharges (Gavaret et al. 2014). The MEG system was a CTF system with 151 z-gradiometers and EEG was recorded using a 64 channel MEG-compatible system. The cortical surface (grey/white matter interface) was segmented with the Brainvisa software (Cointepas et al. 2010). Sensor location and cortical surface were imported into the BrainStorm toolbox (Tadel et al. 2011) within Matlab (Mathworks, Naticks, MA), and the cortical surface was then downsampled to 15004 vertices. The resolution of the EEG/MEG forward model provided the gain matrix linking the amplitude of each cortical dipole, located at each vertex and normal to the surface, to the EEG/MEG sensors. Forward model was estimated using the Boundary Element Model method implemented within the OpenMEEG (Gramfort et al. 2010) plugin of Brainstorm.

The forward model implemented in OpenMEEG is based on equations that only depend on resistivity of the medium, considering that the quasistatic assumption holds at low frequencies. Details are in Gramfort et al. 2010, a summary follows. For the electric field and potential, this results in the law of electrostatics:

$$\nabla \cdot (\sigma \nabla V) = \nabla \cdot J^p \qquad (1)$$

where σ is the conductivity of the medium, V is the electric potential, and $J^p$ dipolar source distribution within the domain. The boundary condition at the interface between piecewise linear conductive regions (between brain and skull, skull and scalp, scalp and air) is:

$$\sigma \nabla V \cdot n = j \qquad (2)$$

With **j** the normal current and **n** the vector normal to the surface.
The magnetic field has a dependence on both the electric potential and on the current source distribution, as described by the Biot and Savart law:

$$B(r) = \frac{\mu_0}{4\pi} \int (J^p(r') - \sigma \nabla V(r')) \times \frac{r-r'}{\|r-r'\|^3} dr' \qquad (3)$$

This formalism assumes that there is no dependence on frequency for the propagation of electric and magnetic fields. This allows computing two gain matrices $G^{MEG}$ and $G^{EEG}$ that link the amplitude of a dipole on the brain surface to the MEG and EEG measures respectively, which then do not depend on the frequency.

A more generic formulation was proposed by Deghani and colleagues (2010), in which the medium is considered to be linear but not purely resistive. This is expressed by equations (10) and (12) in (Dehghani et al. 2010):

$$\nabla \cdot (\gamma_f \nabla V_f) = \nabla \cdot \mathbf{j}_f^p \qquad (4)$$

$$\nabla \times \left(\gamma_f^{inv} \nabla \times \mathbf{B}_f\right) = \mu \nabla \times (\gamma_f^{inv} \mathbf{j}_f^p) \qquad (5)$$

With a complex valued conductivity $\gamma_f = \sigma_f + i\omega\epsilon_f$, which depends on pulsation ω=2πf and permittivity ε$_f$, and 'inv' denoting the inverse. This formulation that prevents from using a single gain matrix for all frequencies, and would require more detailed modelling of the biophysics of the problem at hand, was not considered in our study.

We considered three layers, scalp, outer skull and inner skull, with skull conductivity set as 1/40 of scalp and brain conductivities. This resulted in two gain matrices: $G^{MEG}$ of size 151 channels x 15004 dipoles and $G^{EEG}$ of size 64 channels x 15004 dipolar sources.

Our goal was to generate realistic simulated datasets exhibiting an interaction between spatial scale and temporal frequencies, with high frequencies recruiting small brain areas, and low frequencies recruiting spatially large areas. The dependence between frequency and extent of active cortex has been observed in animal studies (Destexhe et al. 1999). Our working hypothesis was that such spatio-temporal structure alone would exhibit $1/f$ spectral densities differing between EEG and MEG signals due to different spatial 'blurring' properties. To do so, we used the random parcellation function of Brainstorm, with a fixed number of regions $n_s$ (i.e., number of brain patches at scale *s*), varying the spatial scale s of the parcellation by successive factors of two: s=15, 30, ... 7680.

In a given configuration, we chose the specifics scales to be included in the simulation, as well as the frequency ranges to be attributed to each scale. Three scales of parcellation are illustrated in Figure 3 for two configurations (detailed configurations are listed in Table 1). The contribution of a given region to the MEG sensors was obtained by summing the unit contribution of the dipoles within the region (given by the corresponding columns of the gain matrix). We thus obtained at each scale *s* coarse gain matrices $G_s^{MEG}$ of size 151 channels x $n_s$ and $G_s^{EEG}$ of size 64 channels x $n_s$, with $n_s$ being the number of regions at scale *s*.

*Dimensionality of gain matrix*
In order to quantify the variations of dimensionality of the gain matrix across scales, we computed for each scale a singular value decomposition of the EEG and MEG gain matrices. Importantly, each row was standardized (removal of mean and division by standard deviation) in order to compensate for impact of the depth of the source on channel-level amplitudes. We estimated the subspace dimension by finding the first dimension explaining more than 90%, 95% or 99% of the total variance (computed as the cumulative sum of the singular values normalized by the sum of all singular values).

*Generation of signals*
At each parcellation level, a single random time course was applied to each region, and the time course at sensor level was obtained by multiplying the summed contribution of the dipoles in a given region (obtained from simplified gain matrices $G_s^{MEG}$ and $G_s^{EEG}$) with its corresponding time course.

We chose heuristically the range of spatial scales to be considered, either s = 15 to 480 or s= 30 to 3840. The number of regions at scale *s* is given by

$$n_s = n_1 \cdot 2^{(s-1)} \qquad (6)$$

with $n_1$ number of regions at scale 1.

The time course of each region was generated as a white noise (function *rand* of Matlab) considered to be sampled at a frequency $F_s$ of 1250 Hz, with a length of 30s. Importantly, we standardized each signal (removal of mean and normalized by standard deviation) in order to remove the influence of the filter bandwidth on signal amplitude.

The total band of interest was subdivided logarithmically between two bounds that may vary across configurations (details in table 1). Configurations were designed in order to span a range of possibilities

- Configuration 1: mostly low frequencies (1-100Hz) and larges patches (which mimics non-REM sleep-like behavior with large synchronies and dominant low frequencies)
- Configuration 2: higher frequencies and smaller patches, which simulates the awake state
- Configuration 3: spans a large range of sizes and frequencies
- Configuration 4: only one scale, in order to verify that there is no difference on the spectra between EEG and MEG in that case.

Finally, a random offset of +-20% was applied to the limits of each frequency band across regions at a given scale, in order to avoid systematic effects in the spectrum (e.g., notches at frequency boundaries). Large patches were attributed a signal with low frequency contents, while small patches were matched with a signal with high frequency content (see left column of Figure 3 for illustration). The band limited noise $E_s^i$ for each patch *i* and each scale *s* was obtained by filtering white noise in the frequency band $FB_s$ (with offset) adapted to the scale *s*. We used the function *fftfilt* of the EEGlab toolbox (Delorme and Makeig 2004), which is based on Fourier filtering. For each patch i, we compute the corresponding contribution on EEG and MEG by projecting the random activity to the sensors though the corresponding gain matrix:

$$sig_s^i(j,t) = G_s^i(j)E_s^i(t) \qquad (7)$$

with *i* the index of the patch, $E_s^i$ a white noise filtered in band $FB_s$, *j* a channel index and *t* time. The simulated MEG or EEG data *M(j,t)* was obtained by summing over the scales the contributions from all regions i at a given scale. In this summation, the data from each scale s was scaled by a multiplicative coefficient equal to the $s^{th}$ power of a scaling parameter *c* ranging from 1 to 1.9:

$$M(j,t) = \sum_{s=1}^{s=N} \sum_{i=1}^{i=n_s} c^s sig_s^i(j,t) \tag{8}$$

The signal M is at the sensor level. The parameter c allows a fine tuning of the slope of the resulting 1/f spectrum. It can be seen as representing the 'local' level of synchrony on each patch (Cosandier-Rimélé et al. 2007) as a function of frequency. The range of c was chosen in order to result in a realistic range of spectrum slopes.

*Computation of power spectra*

We computed the spectrum S(f) of each channel of the simulated MEG and EEG data using the Welch estimator (*spectrogram* function of Matlab). Thus, the FFT was computed with a window size of 1s (resulting in a frequency resolution of 1Hz), a Hanning tapering, and an overlap of 50% between consecutive windows. The mean spectra across channels were then computed separately for EEG and MEG, and normalized with respect to the higher frequency value, as suggested in (Dehghani et al. 2010). This normalization was necessary in order to compare the shape of the spectra independently of the actual values (differing between EEG and MEG).

If we assume that in a given frequency range the spectrum has a $1/f^\alpha$ shape, then a simple linear regression in a log-log representation will provide an estimation of α:

$$log_{10}(S) = -\alpha\, log_{10}(f) \tag{9}$$

The slopes α of the spectra were estimated with linear regression and compared to one another with a t-test.

**Results**

Figure 4a and 4b present the spectra obtained in configuration 1 (low frequencies configuration) and 2 (high frequencies configuration) respectively. In both cases, the slope between EEG and MEG differ, with MEG presenting less attenuation with increasing frequencies that EEG as hypothesized (i.e., a smaller α in (9)). For configuration 1, EEG slope was -1.55 and MEG slope -1.27, and in configuration 2, EEG slope was -1.73 and MEG slope -1.47.

Figure 4c, corresponds to configuration 3, where we tested a large range of scales, which permits to see that above a particular scale, corresponding to 480 regions, the difference between EEG and MEG spectra seem to disappear.

In Figure 4d, we show the results of configuration 4, with only one scale. As expected, in this minimal space/frequency structure, there is no difference between EG and MEG spectra.

In Figure 5, we illustrate the impact of the scaling parameter c on the slope of the spectra, varying the values around the 'absence of scaling factor' (c=1, i.e. same amplitude at each scale). For a small c (0.6), we obtain large slopes (around -3) and a smaller difference between the EEG and MEG spectra than with c=1. Conversely, with a high value of c (1.4), we obtain smaller slopes and large differences between EEG and MEG

We show in Figure 6 the dimensionality of the gain matrix, as a function of the spatial scale (i.e; number of regions). For a percentage of explained variance of 90%, i.e. at a coarse level of description, EEG and MEG exhibited very similar profiles of dimensionality as a function of number of regions involved in the simulations. For higher levels, i.e. when considering a 'finer' explanation of data at sensor level, the dimensionality of the EEG and MEG behave differently. In particular, we found that EEG dimensionality "saturates" for a number of regions above 480. For 90% of explained variance, there was no notable difference between EEG and MEG. For higher levels, and in particular when accounting for 99% of explained variance, the dimensionality of MEG was higher than EEG, suggesting that it can capture finer spatial details than EEG.

**Discussion**

Several mechanisms have been proposed for explaining the 1/f spectrum in MEG-EEG data. Some arise from the structure of the spatio-temporal dynamics. For instance, Jirsa demonstrated that spatially continuous dynamic systems with homogeneous (gray matter) connectivity always generate $1/f^\alpha$ characteristics in the power spectrum. (Jirsa 2009). Other are related to the non-resistive aspect of the tissues, which lead to filtering effects (Dehghani et al. 2010; see review in Pesaran et al. 2018).

In the present study, we implemented a model with scale-free properties both in the time domain (tuned with a scaling factor *c*) and in the spatial domain (patch size that differ by a factor of two across different scales). This model has both phenomenological aspects, as we used filtered noise, in contrast to computational models, and mechanistic aspects, by relying on a specific space/frequency structure. Importantly, we explicitly introduced a dependence between the size of the patch (spatial scale) and the frequency range and bandwidth it is endowed with: the smaller the patch, the higher the frequency. The model shows that such dependence resulted in differences between EEG and MEG scale-free spectra, in agreement with previous findings obtained both theoretically and in real data (Dehghani et al. 2010). This shows that a specific space/frequency structure of brain activity could generate self-similarity and can can explain differences between spectra, resulting in different slope of the 1/f processes when comparing EEG and MEG. Such a behavior is not exclusive of other mechanisms that are more likely also play a role on real data (Dehghani et al. 2010; G. Buzsaki et al. 2012), but it is to be noted that this mechanism alone seems sufficient to explain the observed differences between the spectra. The full magnitude of the proposed mechanisms requires a systematic quantification of the various contributions, which has so far not been performed, but remains a future task to be performed.

A key assumption of our model is that the spatial extent of temporally coherent sources decreases and scales with frequency, i.e. that high frequencies tend to be more spatially local, whereas low frequencies tend to involve larger spatial cortical regions. This assumption is physiologically plausible, and was specifically stated in the influential Buszaki book "rhythms of the brain" (G Buzsaki 2006): "Long term observations consistently show that coherence of neuronal activity rapidly decreases as a function of distance at high frequency but deceases less for low frequencies.". An intuitive explanation is that it is easier to synchronize oscillators at a low frequency compared to high frequencies, as a small time delay will have a much stronger influence on phase shift at high frequencies compared to low frequencies. Again, from Buszaki book (p. 122) "when the rhythm is fast, only small groups can follow the beat perfectly because of the limitation of axon conductance and synaptic delays. Slower oscillations spanning numerous axon conduction delay periods, on the other hand, allow the recruitment of very large number of neurons. Thus, the slower the oscillation, the more neurons can participate; hence, the integrated mean field is larger." Differences in local and global processing, and effect of distance on spatial coherence was discussed in (P. L. Nunez 2000).

Still, we have to acknowledge that there are only a few observations on real data that support this phenomenon.

Bullock an McClune have reported data from rodent's brain surface recording (electrocorticography, ECoG), that coherence (on temporal signals) decreased more rapidly with frequency when increasing electrode distance (Bullock and McClune 1989). A study performed in humans by the same group, confirmed these findings (Bullock et al. 1995). Destexhe and colleagues have measured the spatial decay of coherence for two different states (slow wave sleep and wake) that involve different frequency content, on the cortex of cat (Destexhe et al. 1999). They found that on, the correlation between LFP signals decreased from 1 to 0.25 over 7mm, whereas the decrease was only from 1 to approximately 0.7 in slow wave sleep (<1Hz)

In (von Stein and Sarnthein 2000), a relationship was suggested from surface EEG between the extent of functional integration and the synchronization frequency.

However, one has to note that observations in human data are challenging. Firstly, it is necessary to have a sufficient spatial coverage (e.g., high density sampling of ECoG electrodes). Secondly, one needs to deal with the convoluted folding of the cortex, which does not allow recording within sulci and could produce abrupt changes when going from one gyrus to another. Volume conduction issues may take place, which make it difficult to assess where the signal actually come from.

The current progress in multi-electrode arrays, and in particular highly conformable ECoG arrays based on polymer thin sheets (Khodagholy et al. 2011), should allow recording across large surfaces and may assess accurately the space/frequency structure of coherent activity, resulting in estimates of extent of coherent cortex and signal attenuation across scales. Interestingly, recent work based on structural MRI has proposed to describe the anatomical connectivity by the eigenvectors of the Laplacian performed on diffucsion tensor imaging graphs. This results in basic spatial patterns (an equivalent of the Fourier basis set, but in the spatial domain), with harmonics representing different spatial extent. These spatial harmonics could be the physical substrate for oscillators at different frequencies, and would fit nicely into our proposed framework. More work is needed based on electrophysiology data, in order to link structure to function within a scale-free framework (Ciuciu et al. 2014).

It was noted from early work, using simultaneous subdural and scalp EEG recordings, that different frequency contents when comparing direct recordings on the brain surface and scalp recordings could originate from the spatial filtering properties of EEG (Pfurtscheller and Cooper 1975). The authors ruled out impedance effects of the skull, which are similar across the frequency range of interest. They proposed instead that the blurring effect of the skull produces summation of activities at different phases, which is particularly visible at high frequencies where a small temporal jitter between different oscillators causes high phase dispersion and thus high signal cancellation (see schematic illustration in Figure 1). Previous modeling studies, investigating signal space/frequency scales reported that " EEG amplitude in each frequency band can be related to the synchrony of the underlying current sources" (Srinivasan et al. 2007), this topic being presented in details in (P. Nunez and Srinivasan 2005). The influence of the geometry on cancellation effects is investigated in (Ahlfors et al. 2010). It is interesting to note that summation do not occur over an infinite number of oscillators (which would result in perfect cancellation), which means that high frequencies still have a chance to 'express' themselves at the surface, as suggested in (von Ellenrieder et al. 2016).

In this context, it is plausible that different spatial filtering properties of MEG and EEG (with EEG being more 'blurred' than MEG because of higher impact of skull conductivity (and more selective impact of dipolar sources orientations on MEG data) can produce different temporal filtering properties. This is indeed what we observed in our simulated data, resulting in a steeper $1/f$ profile for EEG when compared to MEG. Interestingly, a better sensitivity to (relatively high frequency) gamma-band activity in MEG when compared to EEG has been observed during visual stimulation (Muthukumaraswamy and Singh 2013). These results are compatible with differential $1/f$ profiles in the two modalities.

In our simulation study, the differential effects are lost after a certain level of granularity, i.e. when considering synchronicity over only small patches. This finding is also consistent with our study on dimensionality of gain matrix, which demonstrated that the dimension saturates for small patches sizes, although not as fast in MEG when compared to EEG). Interestingly, this saturation of the effect results in a change of slope on the MEG spectra which could be potential explanations for previous observations of abrupt changes of slopes in the spectrum (Miller et al. 2009).

We have simulated a uniform spatio-temporal structure by using an unique scaling parameter c across the whole brain. However, it has been observed regional differences in alpha. In (Dehghani et al. 2010), three main regions are highlighted, parieto-temporal, vertex and frontal areas. These differences could be reproduced by local variations in frequency to scale mapping, or simply with different c parameter across different brain regions.

Finally, it is worth to note that the real-life MEG spectra are usually contaminated by ambient noise, which we do not model in the present study. The study by Deghani and colleagues (Dehghani et al. 2010) , based on empty room recordings, gives an indication of the correction factor that can be applied to the slopes of spectra in order to take into account the contribution of the ambient noise. In this later study, it is shown that the correction will lower the alpha band and thus would amplify the differences between EEG and MEG described in the present work.

**Conclusion**

In conclusion, we demonstrated that the combination of the spatio-temporal structure of brain activity and biophysical effects is a putative mechanism to explain differences in $1/f$ spectra between EEG and MEG. Thus, our simulations provide a phenomenological explanation to previous experimental findings. This is relevant for signal processing, where $1/f$ properties of EEG or MEG signals need to be properly accounted for (Roehri et al. 2016), as well as for recording, as different electrode sizes (resulting in different impedances) integrate brain activity over different volumes (Nelson and Pouget 2012). The scale-free properties of signals attract increasing interest as new biomarkers of brain state. This goes beyond the classical analysis of frequency peaks in the spectrum, by quantifying the non-oscillatory ("arrhythmic") part of the signals (He 2014), which is reflected in the overall slope of the spectrum, independently of the presence of local peaks. Our findings highlight a putative mechanism that could be important to take into account in the interpretation of results in MEG and EEG. In particular, they would need to be corrected for if one wants to integrate EEG and MEG into a common view of scale-free properties of brain signals. Our findings may also potentially have a strong impact on computational modeling of brain signals (Sanz Leon et al. 2013; Sanz-Leon et al. 2015). Indeed, our results suggest that it could be interesting to model the space/frequency properties of neuronal activity, with different local connectivity matrices at different frequencies, which could for example be based on decomposition of tractography-based connectivity matrices (Atasoy et al. 2016).

The processes that we describe here are not exclusive of other possible mechanisms. More investigation should be conducted in evaluation the range of activated cortex as a function of frequency, in varying brain states (in particular awake versus sleep). Future work could address the different contributions of these mechanisms to the observed signals, as well as improvement in biophysical modelling (Bangera et al. 2010). The study of the relationships between structure and function, which is currently attracting much interest, would be particularly relevant in this context (Atasoy et al. 2016; Wirsich et al. 2016).

## Acknowledgments

CGB thanks Jean Gotman for useful discussions on spatial coherence. Research supported by grants ANR-16-CONV-0002 (ILCB), ANR-11-LABX-0036 (BLRI) and ANR-11-IDEX-0001-02 (A*MIDEX)". This work has been carried out within the FHU EPINEXT with the support of the A*MIDEX project (ANR-11-IDEX-0001-02) funded by the "Investissements d'Avenir" French Governement program managed by the French National Research Agency (ANR).Part of this work was funded by a joint Agence Nationale de la Recherche (ANR) and Direction Génerale de l'Offre de Santé (DGOS) under grant "VIBRATIONS" ANR-13-PRTS-0011-01. This work was performed within a platform member of France Life Imaging network (grant ANR-11-INBS-0006)

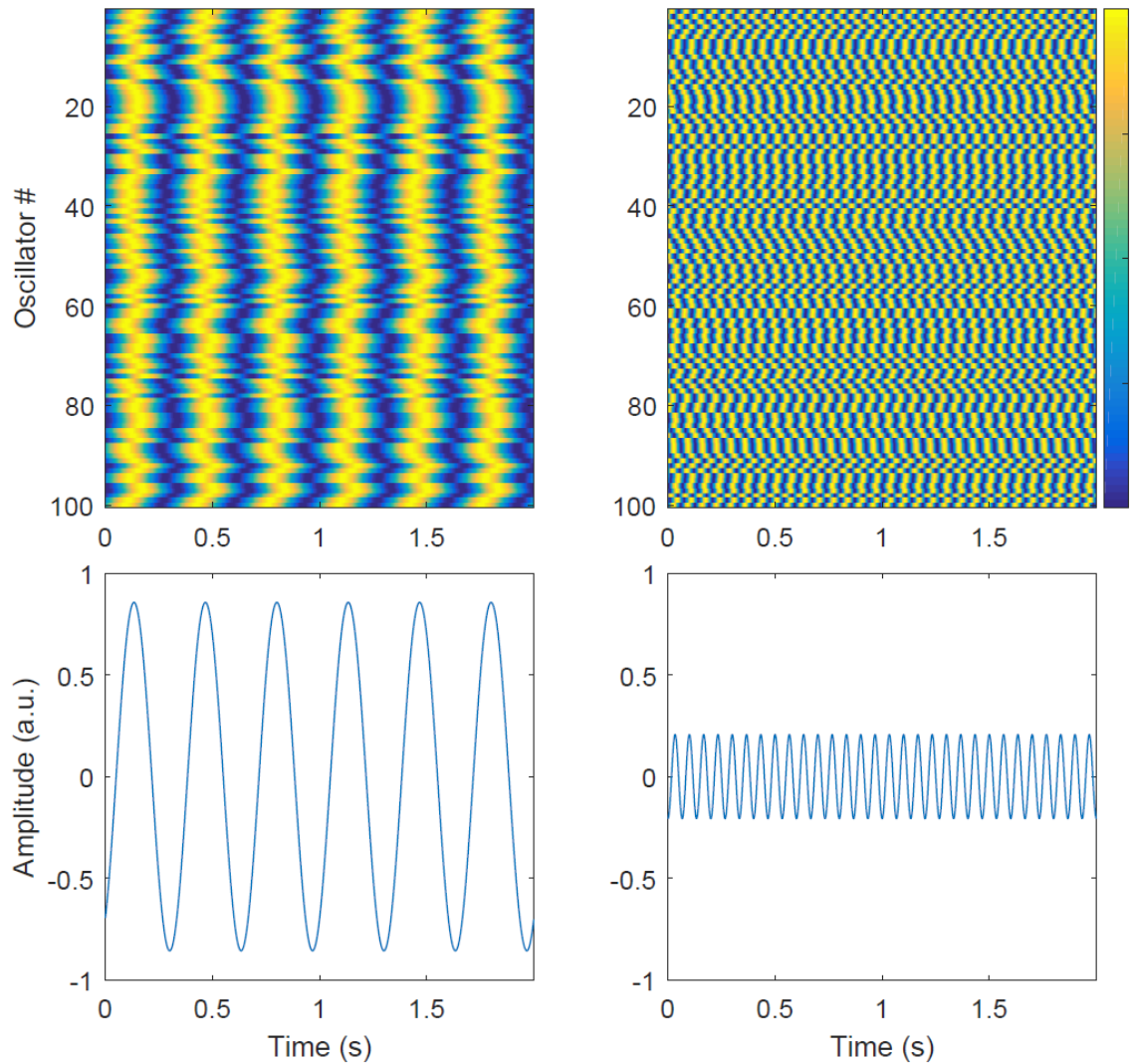

Figure 1: Schematic illustration of differential effect of spatial smoothing at different frequencies. The top row shows realizations of a sine wave with random jitter (taken uniformly in 0-100ms). These signals can be seen as representing neighboring neuronal oscillators from a small brain region. The bottom row is the average across realizations. The cancellation resulting from averaging is smaller for the low frequency (left column, 3 Hz oscillations) than for high frequency (right column, 15 Hz oscillations). This effect was discussed in (Pfurtscheller and Cooper 1975).

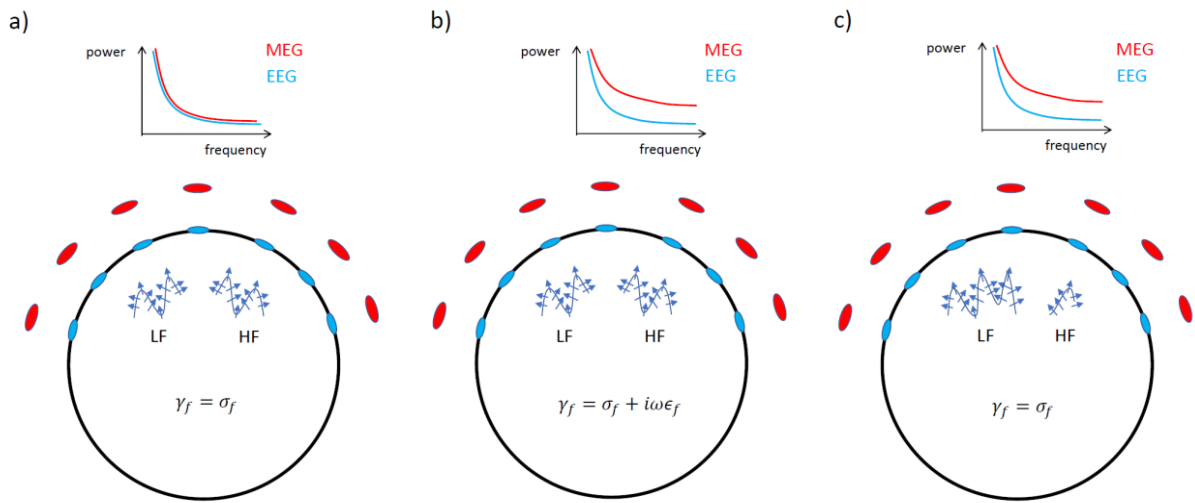

Figure 2: Schematic view of different biophysical scenarios for explaining differences between EEG and MEG spectra. **a)** the conductive medium is purely resistive (i.e., the electric potential follows Ohm's law U= σ I with U potential, σ resistance and I intensity), and propagation of electric and magnetic does not depend on frequency. Two schematic patches of cortex are represented, with synchronous activity involving equal surfaces, one for the low frequencies (LF) and one for high frequencies. The arrows represent equivalent current dipoles corresponding to the activity of pyramidal cells. Here, the spectra for EEG and MEG are the same. **b)** The patches of active cortex remain of equal size for high and low frequencies, but the medium is not purely resistive, i.e. propagation of electromagnetic fields depend on frequency (as proposed by Deghani et al 2010). Here, EEG and MEG spectra differ due to the properties of the conductive medium. c) Low frequencies involve large patches of cortex, while high frequencies involve synchrony over smaller regions. Here, the spectra also differ, but this time because of the different smoothing properties of EEG and MEG, as EEG integrates activity over larger volumes. This is the hypothesis proposed in the current study.

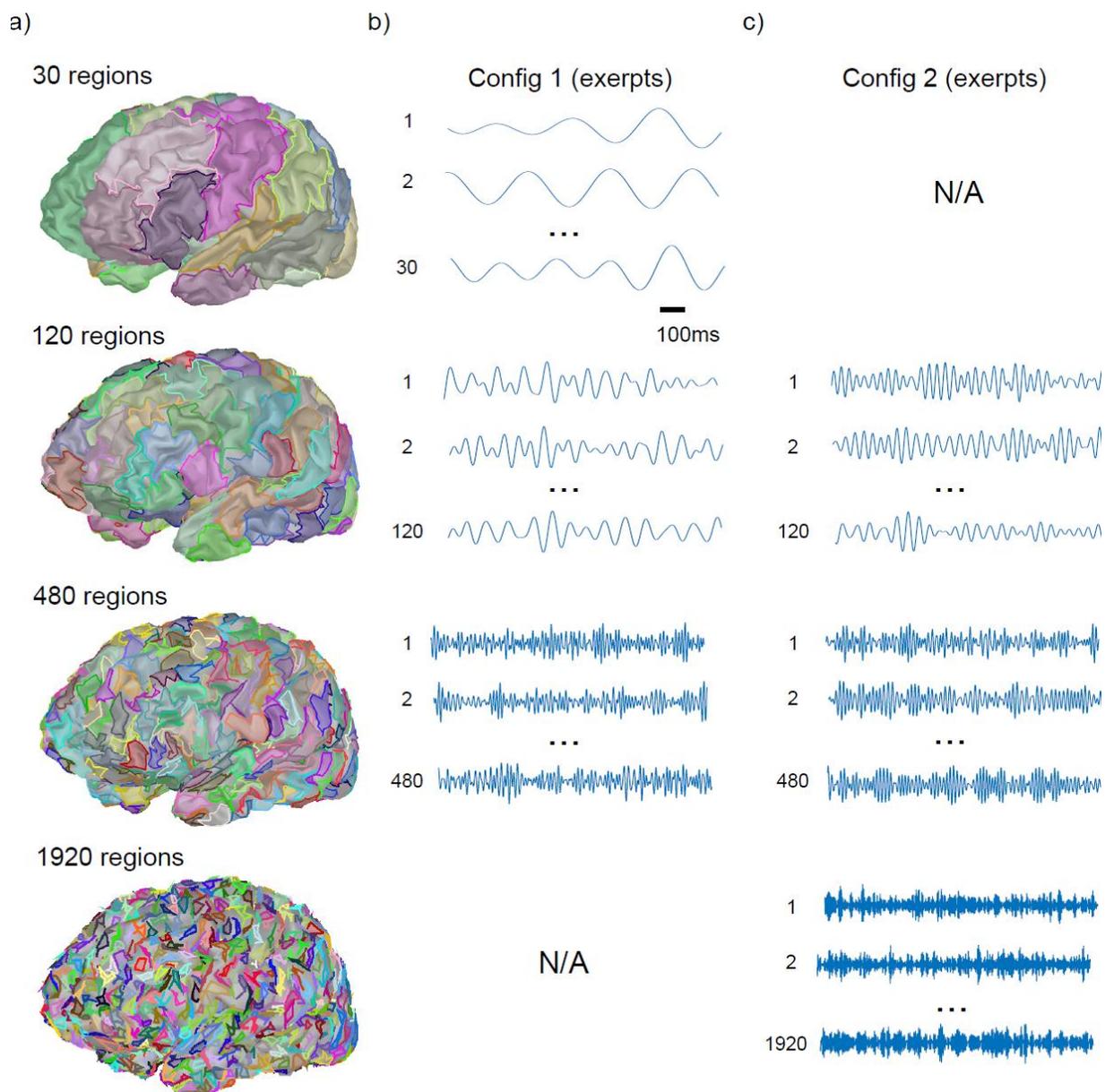

Figure 3: Illustration of the simulation framework. a) Examples of levels of brain parcellation. At each level, a single time course is applied to all the dipoles within a given region. The bandwidth of the random noise that is applied to each region varies with the patch size, low frequencies for large regions and high frequencies for small regions (see Table a) b) Signals corresponding to configuration 1 (low frequencies, only three scales are shown here) c) Signals corresponding to Configuration 2 (high frequencies, only three scales are shown here) The resulting data for each configuration is the sum of the simulated MEG signals generated across levels.

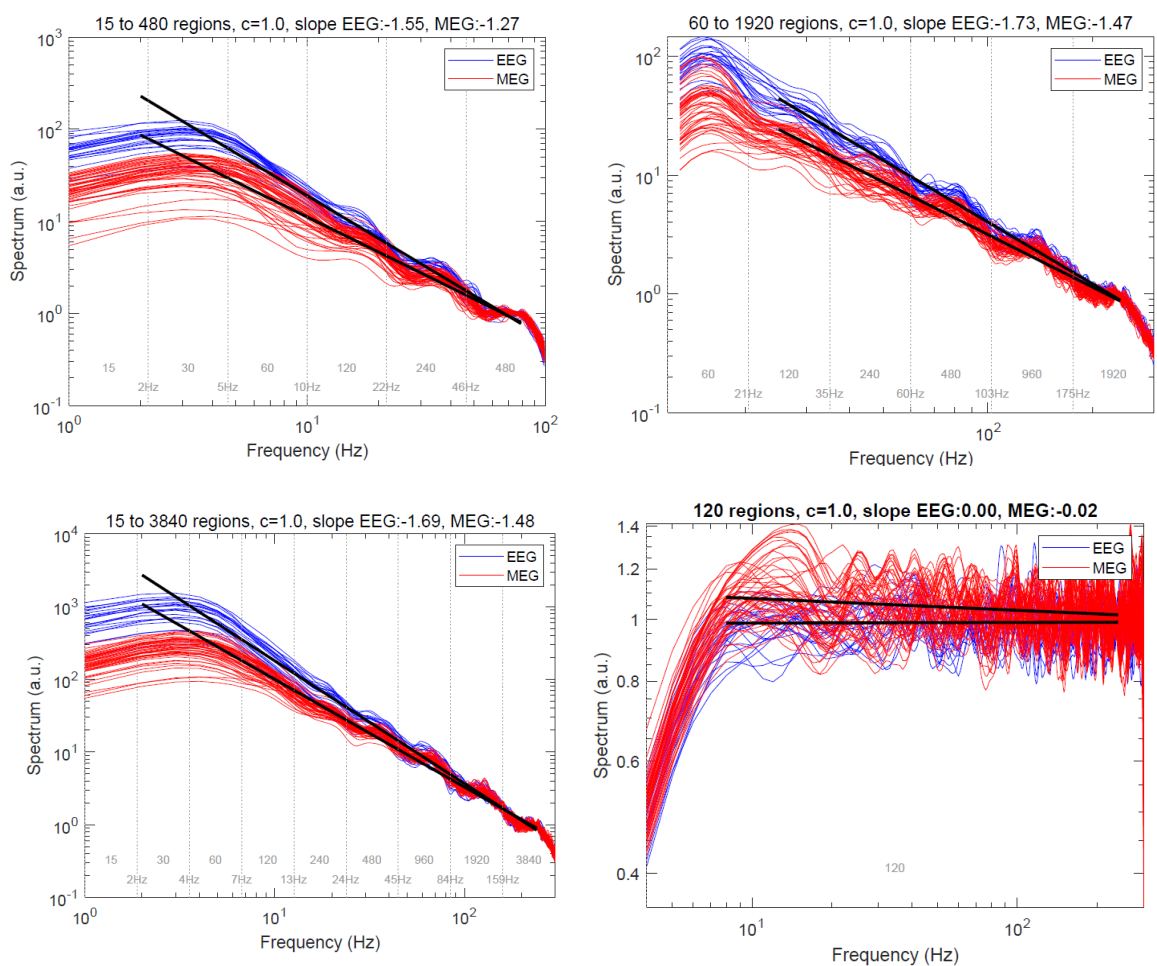

Figure 4: EEG and MEG spectra obtained for the different configuration, with a scale parameter c=1 (only 1/3 of channels are shown). a) Configuration 1 (low frequencies) b) Configuration 2 (high frequencies) c) Configuration 3 (large range of scales) d) Configuration 4 (only one scale). All the configuration presenting a space/frequency structure (1-3), there is a difference between EEG and MEG spectra, at least below a certain scale (corresponding to 480 regions).

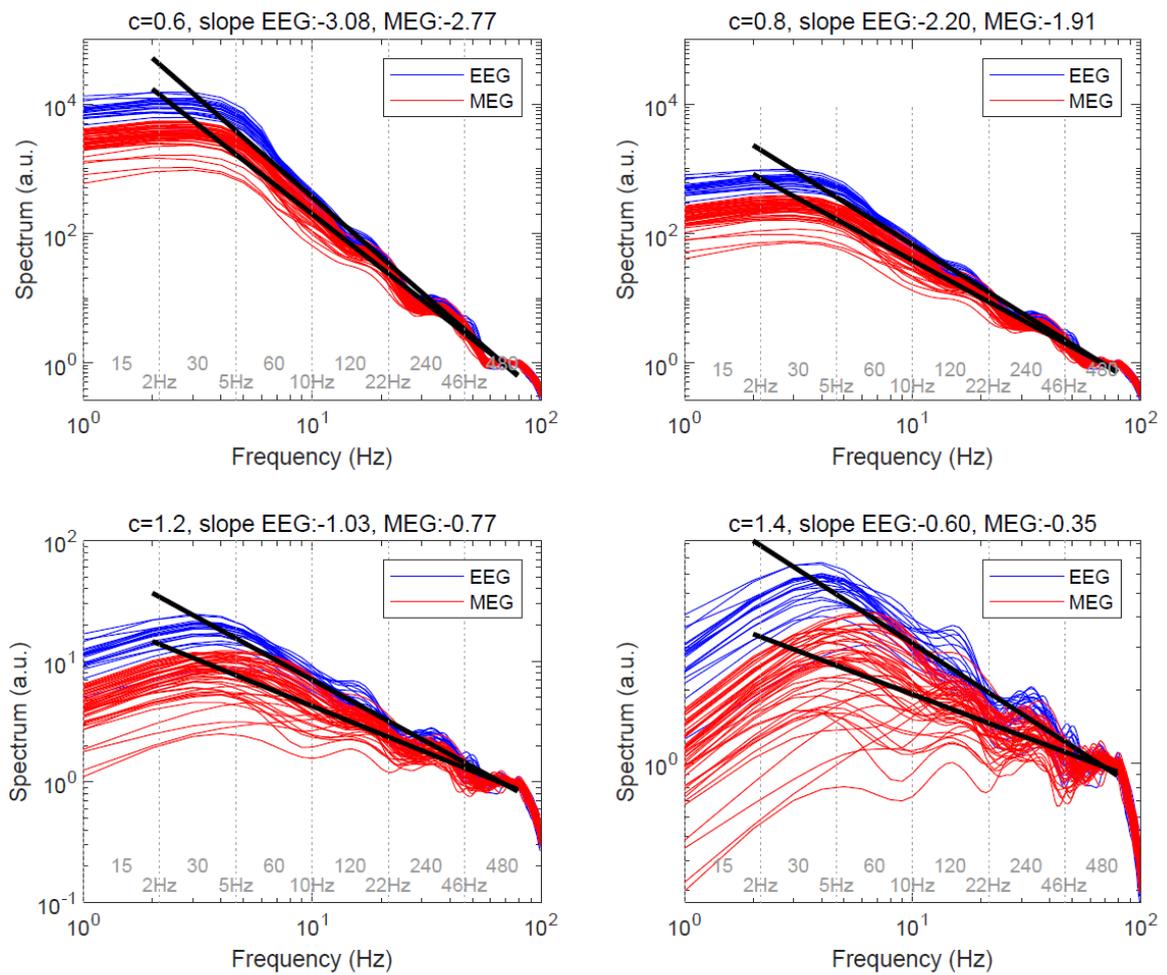

Figure 5: Comparison of the slopes of EEG and MEG average spectra across sensors, in Configuration 1 (see Table 1), for different parameters of the amplitude scaling coefficient *c*. The absolute value of the slope increases with smaller c, as expected, with alpha ranging between 3.08 and 0.35. Differences between EEG and MEG are higher for higher c.

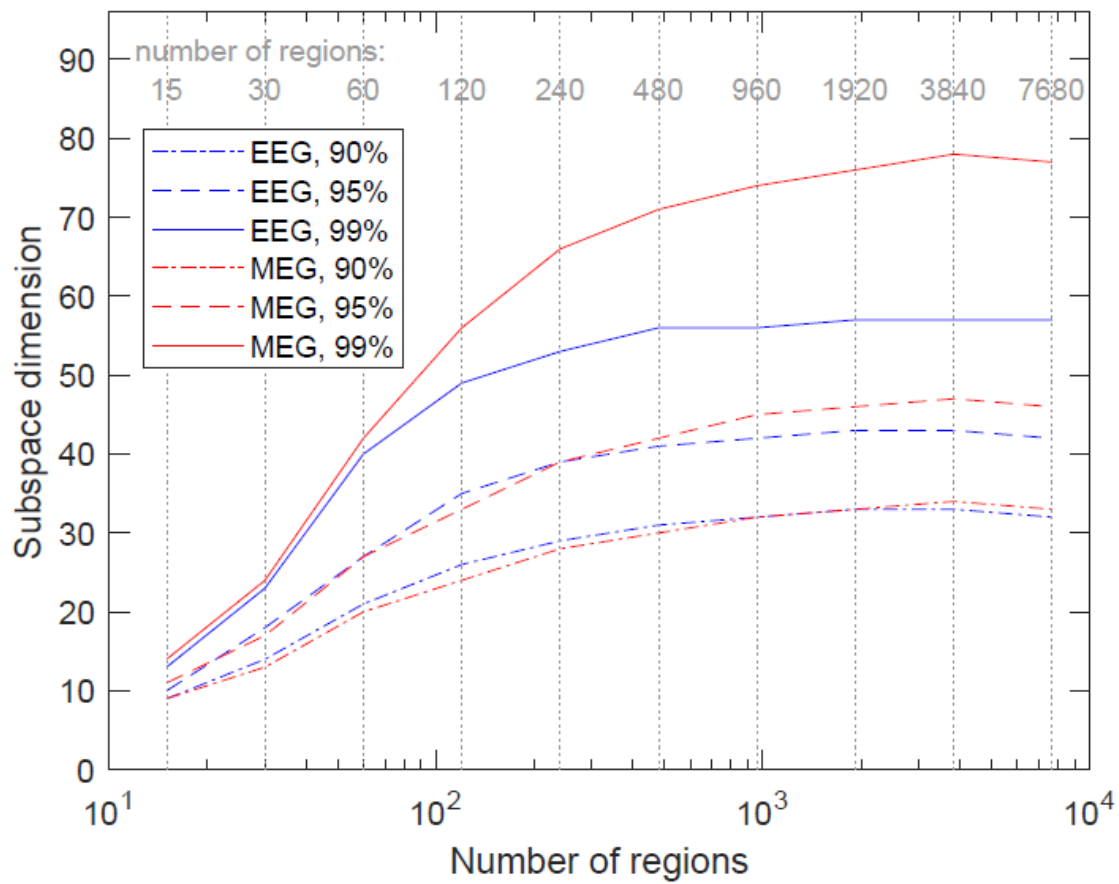

Figure 6: Dimensionality analysis of the EEG and MEG gain matrices as a function of the spatial scale (number of regions), for different values of explained variance (90%, 95%, 99%). For 90% of explained variance, there was no notable difference between EEG and MEG. For higher levels, and in particular when accounting for 99% of explained variance, the dimensionality of MEG was higher than EEG, suggesting that it can capture finer spatial details than EEG.